\documentclass[letter]{aa}
\usepackage{lscape}
\usepackage{graphicx}

\begin{document}


\title {Chemical differentiation in a prestellar core traces non-uniform illumination \thanks{Based on observations carried out with the IRAM 30m Telescope. IRAM is supported by INSU/CNRS (France), MPG (Germany) and IGN (Spain)}
}


\author{S. Spezzano\inst{1} \and L. Bizzocchi\inst{1} \and P. Caselli\inst{1} \and J. Harju\inst{1,2} \and S. Br{\"u}nken\inst{3}}
 \institute{Max Planck Institute for Extraterrestrial Physics, Giessenbachstrasse 1, 85748 Garching, Germany \and Department of Physics, P.O. Box 64, University of Helsinki, Finland \and I. Physikalisches Institut, Universit\"at zu
  K\"oln, Z\"ulpicher Str. 77, 50937 K\"oln, Germany }


\abstract
{Dense cloud cores present chemical differentiation due to the different distribution of C-bearing and N-bearing molecules, the latter being less affected by freeze-out onto dust grains. In this letter we show that two C-bearing molecules, CH$_3$OH and  $c$-C$_3$H$_2$, present a strikingly different (complementary) morphology while showing the same kinematics toward the prestellar core L1544.  After comparing their distribution with  large scale H$_2$ column density $N$(H$_2$) map from the {\em Herschel} satellite, we find that these two molecules trace different environmental conditions in the surrounding of L1544: the $c$-C$_3$H$_2$ distribution peaks close to the southern part of the core, where the surrounding molecular cloud has a $N$(H$_2$) sharp edge, while CH$_3$OH mainly traces the northern part of the core, where $N$(H$_2$) presents a shallower tail. We conclude that this is evidence of chemical differentiation driven by different amount of illumination from the interstellar radiation field: in the South, photochemistry maintains more C atoms in the gas phase allowing carbon chain (such as $c$-C$_3$H$_2$) production; in the North, C is mainly locked in CO and methanol traces the zone where CO starts to freeze out significantly. During the process of cloud contraction, different gas and ice compositions are thus expected to mix toward the central regions of the core, where a potential Solar-type system will form. An alternative view on carbon-chain chemistry in star-forming regions is also provided.}

   \maketitle

\section{Introduction}
Prestellar cores represent the initial conditions of star formation and they are precious laboratories to study
physical and chemical processes away from the complications due to protostellar feedback. Their chemical composition provides the starting point out of which the future protoplanetary disk and stellar system will form.  Molecular studies also provide important clues on the dynamical status and evolution. Based on spectroscopic data, \cite{ket15} in fact concluded that the prestellar core L1544 is slowly contracting and its kinematics is not consistent with the singular isothermal sphere \citep{shu77} and Larson-Penston \citep{lar69, pen69} contraction.  \cite{cas99, cas02a} showed that CO is heavily ($>$99\%) frozen onto the surface of dust grains in the central 6500\,AU of L1544, while N$_2$H$^+$ and, in particular N$_2$D$^+$, better follow the millimetre dust continuum emission showing no clear signs of depletion (see also Crapsi et al. 2007 for similar conclusions on NH$_2$D). \cite{taf02, taf04, taf06} extended this study to more starless cores and more molecular species. They found a systematic molecular differentiation, pointing out that C-bearing species, such as CO, CS, C$_2$S, CH$_3$OH, C$_3$H$_2$, and HC$_3$N are severely affected by freeze-out, showing a sharp central hole (with some differences in hole size depending on the particular molecule), while N$_2$H$^+$ and NH$_3$ seem present in the gas phase at the core centres. Carbon chain molecules are known to preferentially trace starless and less evolved cores \citep{suz92}, where C atoms have not yet been mainly locked in CO. CCS has also been found in the outer outer edge of L1544 \citep{oha99} and, in the case of L1551, with clear shift in velocity compared to the LSR core velocity \citep{swi06}, indicating possible accretion of material from the more diffuse molecular cloud where the dense core is embedded. Although gas-grain chemical models (of spherically symmetric starless cores) can explain the observed differential distribution of C-bearing and N-bearing molecules based on the chemical/dynamical evolution (e.g. Aikawa et al. 2001), observational and theoretical studies have so far not discussed possible differences in the distribution of C-bearing molecules affected by central freeze-out.  Here we report on the distribution of methanol (CH$_3$OH ) and cyclopropenylidene ($c$-C$_3$H$_2$) toward L1544 and show that their complementary distribution can be understood if the non-homogeneous environmental conditions are taken into account. 

The recent CH$_3$OH map of \cite{biz14} shows an asymmetric structure, consistent with central depletion and preferential release in the region where CO starts to significantly freeze out (CH$_3$OH is produced on the surface of dust grains via successive hydrogenation of CO; e.g. Watanabe $\&$ Kouchi 2002).  In particular, the CH$_3$OH column density map presents a clear peak toward the North of the millimetre dust continuum peak (see Figure 1 of Bizzocchi et al. 2014), thus revealing a non-uniform distribution around the ring-like structure. With the aim of exploring chemical processes across L1544, we mapped the emission of another carbon-bearing molecule, $c$-C$_3$H$_2$,  representative of carbon-chain chemistry \citep{sak08}. In this paper we investigate the effects of physical parameters on the distribution of methanol and cyclopropenylidene in L1544, and the influence of environmental effects on the gas- and grain-phase chemistry.

\section{Observations}
\subsection{Cyclopropenylidene}
The emission map of cyclopropenylidene towards L1544 has been obtained in October 2013 using the IRAM 30m telescope (Pico Veleta, Spain), during the same observing run in which methanol was observed \citep{biz14}.
We performed a 2.5\arcmin $\times$2.5\arcmin on-the-fly (OTF) map centred on the source dust emission peak ($\alpha _{2000}$ = 05$^h$04$^m$17$^s$.21,  $\delta _{2000}$ = +25$^\circ$10$'$42$''$.8). The reference position was set at (-180\arcsec,180\arcsec) offset with respect to the map centre. The observed transitions are summarised in Table \ref{table:parameters}. The CH$_3$OH observations are described in \cite{biz14}. An additional setup was used to observe one line of $c$-C$_3$H$_2$. The EMIR E090 receiver was used with the Fourier Transform Spectrometer backend (FTS) with a spectral resolution of 50 kHz. The antenna moved along an orthogonal pattern of linear paths separated by 8$''$ intervals, corresponding to roughly one third of the beam FWHM. The mapping was carried out in good weather conditions ($\tau$ $\sim$ 0.03) and a typical system temperature of T$_{sys}$ $\sim$ 90 K. The data processing was done using the GILDAS software \citep{pet05}, and CASA \citep{CASA}.

\subsection{H$_2$}
The Herschel/SPIRE images were extracted from the pipeline-reduced
images of the Taurus complex made in the course of the Herschel
Gould Belt Survey \citep{Gould}. The data were downloaded
from the Herschel Science Archive
(HSA)\footnote{www.cosmos.esa.int/web/herschel/science-archive}. We
calculated the $N$(H$_2$) using only the
three SPIRE \citep{SPIRE} bands at $250\,\mu$m,
$350\,\mu$m, and $500\,\mu$m, for which the pipeline reduction
includes zero-level corrections based on comparison with the Planck
satellite data. A modified blackbody function with the dust emissivity
spectral index $\beta=1.5$ was fitted to each pixel, after smoothing the
$250\,\mu$m and $350\,\mu$m images to the resolution of the
$500\,\mu$m image ($\sim 40\arcsec$), and resampling all images to the
same grid. For the dust emissivity coefficient we
adopted the value from \cite{Beta}, $\kappa_{250\mu
  m}=0.1$ cm$^{2}$g$^{-1}$. \cite{Beta1} derived a similar value for
$\kappa_{250\mu m}$ in the starless core CrA C. The obtained maximum column density is
$N$(H$_2$)=2.8$\times$10$^{22}$ cm$^{-2}$, which is comparable to the value of 9.4$\times$10$^{22}$ cm$^{-2}$ derived in \cite{cra05} at the 1.2 mm dust emission peak defined in \cite{war99}, if we consider the different beams (40\arcsec versus 11\arcsec).

\begin{table}
\caption{Spectroscopic parameters of the observed lines\tablefootmark{a}}
\label{table:parameters}
\scalebox{0.8}{
\begin{tabular}{ccccc}
\hline\hline
Molecule & Transition & Rest frequency & E$_{up}$& A\\
&  $J'_{K_a',K_c'} - J''_{K_a'',K_c''}$    &(MHz)   & (K)  & ($\times$10$^{-5}$ s$^{-1}$)\\
\hline
$c$-C$_3$H$_2$  &$3_{2,2} - 3_{1,3}$    &  84727.688     &  16.1      &  1.04\\
CH$_3$OH   &  2$_{1,2}$-1$_{1,1}$ ($E_2$) &   96739.362       &  12.53\tablefootmark{b}   & 0.26\\
\hline
\end{tabular}
}
\tablefoot{
\tablefoottext{a}{The map of methanol in L1544 is already presented in \cite{biz14}}
\tablefoottext{b}{Energy relative to the ground 0$_{0,0}$, $A$ rotational state.}
}
\end{table}

\begin{figure}
 \centering
 \includegraphics [width=0.45\textwidth]{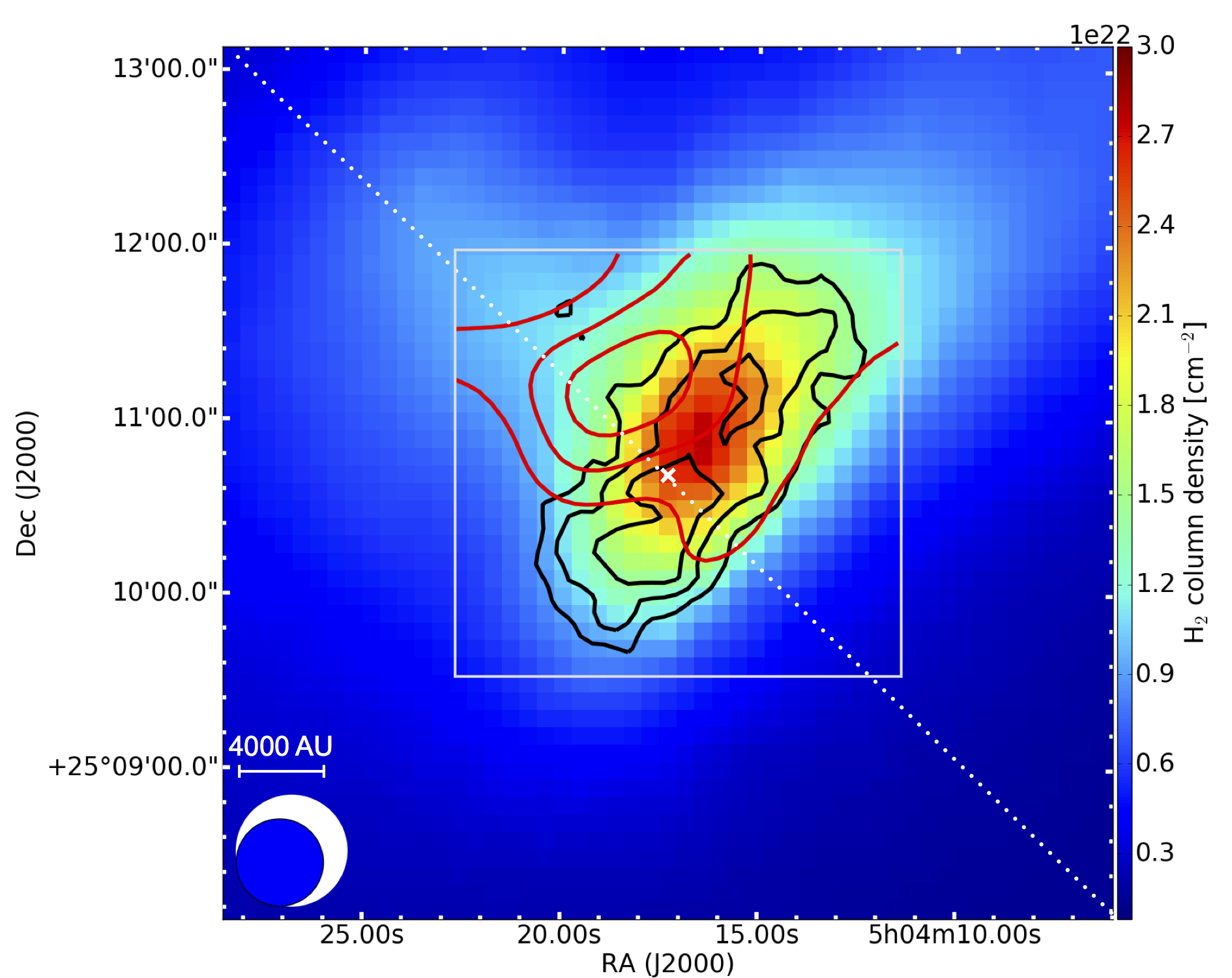}
 \caption{Integrated intensities of the 2$_{1,2}$-1$_{1,1}$ ($E_2$) transition of methanol (red contour) and 3$_{2,2}$-3$_{1,3}$ transition of cyclopropenylidene (black contour) are plotted on the H$_2$ column density map derived from far-infrared images observed by {\sl Herschel} \citep{Hershel}. The contour levels are 10$\sigma$, 15$\sigma$ and 20$\sigma$ CH$_3$OH (2$\times$10$^{-1}$, 3$\times$10$^{-1}$ and 4$\times$10$^{-1}$ K\,km\,s$^{-1}$, respectively) and 5$\sigma$, 10$\sigma$ and 15$\sigma$ $c$-C$_3$H$_2$ (4.8$\times$10$^{-2}$, 7.2$\times$10$^{-2}$ and 9.6$\times$10$^{-2}$ K\,km\,s$^{-1}$, respectively). The white box defines the area that was mapped with the 30m telescope. The white cross marks the position of the 1.2 mm dust emission peak from \cite{war99}. The beams of SPIRE and 30m telescope are shown as a white and blue circle, respectively. The dotted line represents the positions where the column densities plotted in Figure \ref{fig:plot} have been extracted. Methanol and cyclopropenylidene clearly trace different regions in L1544, and show a chemical differentiation within the prestellar core.}
\label{fig:methanol}
\end{figure}

\section{Results and Discussion}
\subsection{Chemical differentiation in L1544}
In Figure \ref{fig:methanol}, the integrated intensities of both the 2$_{1,2}$-1$_{1,1}$ ($E_2$) transition of methanol from \cite{biz14} (red contours), and the 3$_{2,2}$-3$_{1,3}$ transition of cyclopropenylidene (black contours), are superimposed on the H$_2$ column density derived from far-infrared images observed by Herschel \citep{Hershel}. The white box defines the area mapped with the 30m telescope.  

The $N$(H$_2$) map shows a sharp and straight edge toward the South and South-West part of the cloud, which marks the edge of the filamentary cloud within which L1544 is embedded (see also Tafalla et al. 1998). Therefore, 
this side of the dense core should be more affected by the photochemistry activated by the interstellar radiation field (ISRF). In fact, 
methanol shows a complementary distribution with respect to cyclopropenylidene. Is this expected based on our understanding of the chemistry of these two molecules?

\begin{figure}
 \centering
 \includegraphics [width=0.45\textwidth]{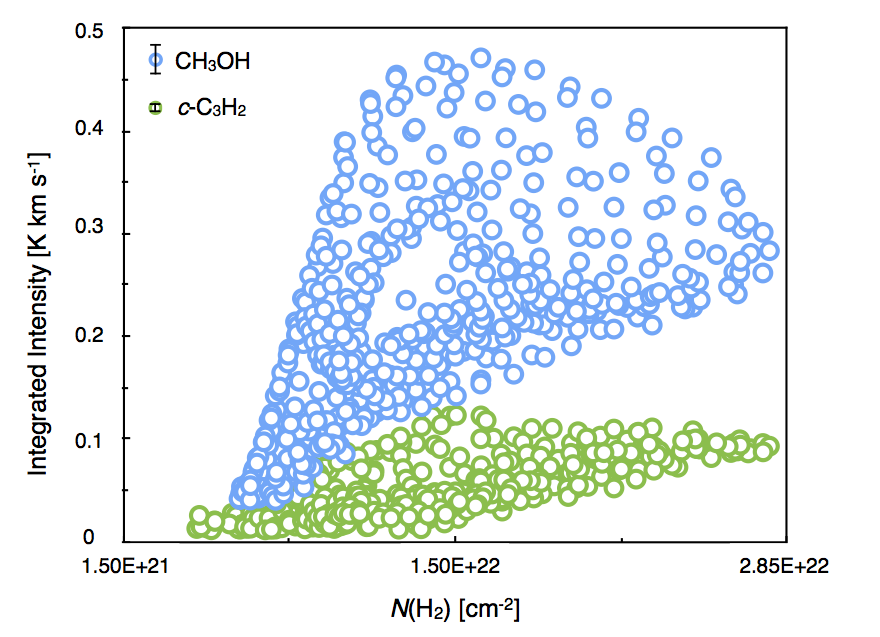}
 \caption{Scatter plot of the integrated intensities of the 2$_{1,2}$-1$_{1,1}$ ($E_2$) CH$_3$OH line (blue circles) and the 3$_{2,2}$-3$_{1,3}$ $c$-C$_3$H$_2$ line (green circles) with respect to the H$_2$ column density inferred from the SPIRE observations (see text). The average error bars are shown on the upper part of the plot. The average errors are 14 mK km s$^{-1}$ for CH$_3$OH and 4 mK km s$^{-1}$ for $c$-C$_3$H$_2$. Only the pixels where the integrated intensity over the average error is larger than 3 are plotted. }
  \label{fig:scatter}
\end{figure}

\begin{figure}
 \centering
 \includegraphics [width=0.45\textwidth]{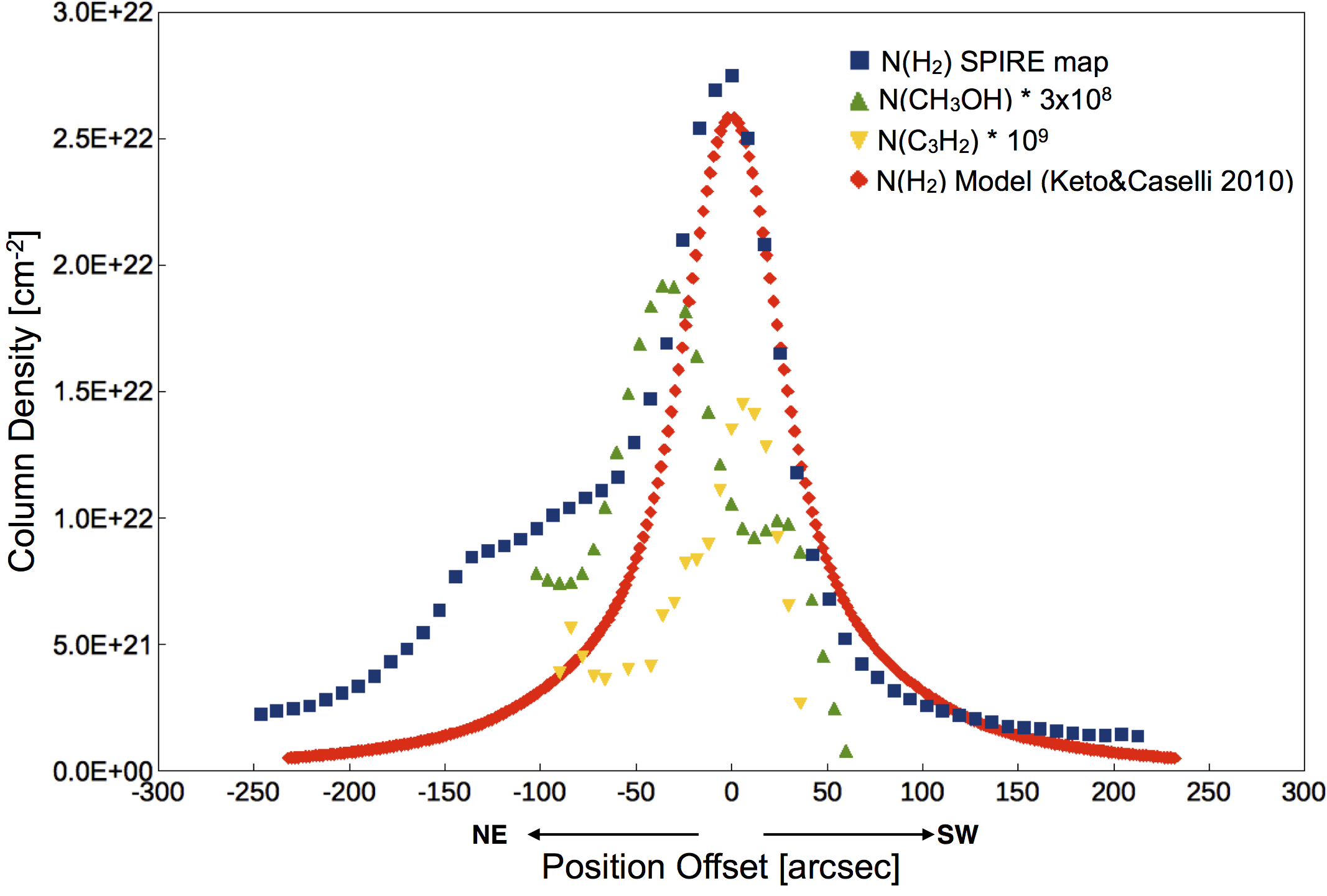}
 \caption{Column densities of H$_2$, $c$-C$_3$H$_2$ and CH$_3$OH extracted along the dotted line present in Figure \ref{fig:methanol}, as well as the $N$(H$_2$) calculated with the model of L1544 described in \cite{ket10} smoothed at 40\arcsec. While the $N$(H$_2$) presents a sharp drop towards the South-West, its decrease is not as steep towards the North-East. The resulting different illumination on the two sides of L1544 is likely to cause the different distribution of cyclopropenylidene and methanol within the core. }
\label{fig:plot}
\end{figure}

Methanol is believed to be formed on dust grains \citep{wat02} by subsequent hydrogenation of carbon monoxide, and its detection towards prestellar cores is already a challenge for current models given the absence of efficient desorption processes in these sources. Thermal desorption is out of question, because of the low dust temperature. Recent laboratory studies show that also the photo-desorption of methanol from ices is negligible \citep{ber16,cru16}. The main desorption products when irradiating pure and mixed methanol ices are photo-fragments of methanol. An alternative route to explain the presence of methanol in the gas phase is the reactive/chemical desorption that has been theoretically proposed by \cite{gar07} and \cite{vas13} and experimentally studied by \cite{dul13} and \cite{min16}. On the other hand, $c$-C$_3$H$_2$ is mainly formed in the gas phase (e.g. Spezzano et al. 2013) and it should preferentially trace dense and 'chemically young' gas, i.e. gas where C atoms have not yet been mainly locked into CO. This C-atom rich gas is expected in the outer envelope of an externally illuminated dense core (e.g. Aikawa et al. 2001). However, toward L1544, $c$-C$_3$H$_2$ only appears to trace one side of the core, the one closer to the sharp $N$(H$_2$) edge and away from the CH$_3$OH peak. This indicates that photo-chemistry is not uniformly active around L1544, most likely due to the fact that the distribution of the envelope material (belonging to the filament within which L1544 is embedded), is not uniform as clearly shown by the {\em Herschel} map in Figure 1. 
Figure \ref{fig:methanol} shows that methanol traces a region farther away from the Southern sharp-edge of the $N$(H$_2$) map, possibly more shielded from the ISRF and where most of the carbon is locked in CO. CH$_3$OH is preferentially found at the northern edge of L1544 because here photochemistry does not play a major role (so C is locked in CO) and the density is low enough to maintain a higher fraction of CH$_3$OH in the gas phase, but above the threshold value for CO freeze out, a few $\times$ 10$^4$ cm$^{-3}$ \citep{cas99}. Based on the \cite{ket10} model (and updated by Keto et al. 2014), the volume density at the distance of the methanol peak is predicted to be 8$\times$10$^4$ cm$^{-3}$, which is just above the threshold value.
On the contrary, cyclopropenylidene has the most prominent peak toward the Southern sharp-edge of the H$_2$ column density and extends along the semi-major axis of the core, almost parallel to the South-West edge of the $N$(H$_2$) map.
This behaviour is also clearly shown in Figure \ref{fig:scatter} where the integrated intensities of both methanol and cyclopropenylidene are plotted against $N$(H$_2$). $c$-C$_3$H$_2$ is in fact present also at lower $N$(H$_2$) values with respect to methanol, and it maintains a flat intensity profile, suggestive of a layer-like structure, with no significant increase toward the core center. CH$_3$OH intensity instead shows a sharp rise up to column densities of about 1.6$\times$10$^{22}$ cm$^{-2}$ and it declines at higher values.

Figure \ref{fig:plot} shows the column densities of H$_2$, $c$-C$_3$H$_2$ and CH$_3$OH extracted along the dotted line present in Figure \ref{fig:methanol}, as well as the $N$(H$_2$) calculated with the model of L1544 described in \cite{ket10} assuming a beam size of 40\arcsec. The column densities of $c$-C$_3$H$_2$ and CH$_3$OH have been calculated assuming that the lines are optically thin, using the formula given in Equation (1) of \cite{spe16}.  We assumed a T$_{ex}$ of 6 K for $c$-C$_3$H$_2$ and 6.5 K for CH$_3$OH, as done in \cite{spe13} and \cite{biz14} respectively. In the same works it is reported that both lines present moderate optical depths ($\tau$ < 0.4). This plot shows that the decrease of $N$(H$_2$) towards the South-West, where cyclopropenylidene is more abundant, is much steeper than towards the North-East, where methanol is more abundant. A full map of the $N$($c$-C$_3$H$_2$)/$N$(CH$_3$OH) column density ratio can be seen in Figure \ref{fig:ratio}, showing a clear peak toward the South-East side of L1544.
Figure \ref{fig:plot} also shows that both $c$-C$_3$H$_2$ and CH$_3$OH belong to the same dense core (identified by the brightest peak in $N$(H$_2$)), while tracing different parts of it. We obtain the same result by comparing the line-width and velocity maps of both molecules, see \ref{fig:vlsr}. Despite the different spatial distributions, the two molecules trace the same kinematic patterns (velocity gradient and amount of non-thermal motions). This indicates that the velocity fields are dominated by the bulk motions (gravitational contraction and rotation) of the prestellar core, which similarly affect the two sides of the core and do not depend on the chemical composition of the gas. In summary, both $c$-C$_3$H$_2$ and CH$_3$OH trace different parts of the same dense core and no velocity shift is present (unlike for the L1551 case; Swift et al. 2006). We note that the observed transitions of $c$-C$_3$H$_2$ and CH$_3$OH have relatively high critical densities (between a few $\times$ 10$^4$ and 10$^6$\,cm$^{-3}$), so that these lines are not expected to trace the more diffuse filamentary material surrounding the prestellar core.

\subsection{Carbon-Chain Chemistry}
The above results show that Carbon-Chain Chemistry (CCC) is active in the chemically and dynamically evolved prestellar core L1544 in the direction of its maximum exposure to the ISRF (witnessed by the sharp edge in the H$_2$ column density map; see Fig. 1).  Based on our observations, we conclude that CCC can also be present in slowly contracting clouds such as L1544 (Keto et al. 2015) and that the C atoms driving the CCC can partially deplete on dust grain surfaces, producing solid CH$_4$, which is the origin of the so-called Warm-Carbon-Chain-Chemistry (WCCC; Sakai \& Yamamoto 2013), when it is evaporated in the proximity of newly born young stellar objects. This apparently contradicts the suggestion that WCCC becomes active when the parent core experiences a fast contraction on a time scale close to that of free fall \citep{sak08, sak09}. In these previous papers, fast contraction is invoked to allow C atoms to deplete onto dust grains before they are converted to CO in the gas phase \citep{sak13}, so that a substantial amount of solid CH$_4$ is produced. 

Based on the results of our observations, we propose an alternative view to the WCCC origin, that is not depending on the dynamical evolution: cores embedded in low density (and low A$_v$) environments, where the ISRF maintains a large fraction of the carbon in atomic form in most of the surrounding envelope, become rich in solid CH$_4$ and carbon chains, precursors to WCCC. Cores such as L1544, which are embedded in non uniform clouds, with non uniform amounts of illumination from the ISRF, should have mixed ices, where CH$_4$ and CH$_3$OH ices coexist. It is interesting to note that CH$_3$OH has been recently found to correlate with the C-chain molecule C$_4$H toward embedded protostars \citep{gra16}, suggesting that these two classes of molecules can indeed coexist. We suggest that the sources studied by Graninger et al. (2016) had non-uniform environmental similar to those around L1544, whereas the WCCC sources formed within prestellar cores with most of their envelope affected by photochemistry (due to a lower overall H$_2$ column density or extinction toward their outer edges).  We are now extending our study of CH$_3$OH and $c$-C$_3$H$_2$ to a larger sample of starless cores and link the results to {\em Herschel} $N$(H$_2$) maps, to see if the same conclusion reached toward L1544 can be extended to other regions.

\section{Summary}
We have compared the spatial distribution of $c$-C$_3$H$_2$ and methanol and underlined a chemical differentiation within L1544. While cyclopropenylidene peaks toward the sharp edge of the H$_2$ column density map, where a fraction of the carbon is still in atomic form, methanol has its maximum away from this edge. This suggests that active carbon chemistry driven by free carbon is at work on the South-Western side of L1544, which is exposed to the ISRF. Because of the high density, accretion of carbon atoms onto grains should be effective in this region, and ices may become rich in CH$_4$. On the opposite side, carbon is mostly locked in CO and methanol is formed on the surface of dust grains following CO freeze out.  Here, ices are expected to be CO and methanol rich.  The two sides of the core are however following the same dynamics and it is expected that mixed ices will be present toward the core centre, where a protostar will form. This mixing would provide a natural explanation for the correlation between CH$_3$OH and C$_4$H found recently in a sample of embedded protostars.

\noindent {\it Acknowledgement}\\
The authors wish to thank the anonymous referee for useful suggestions, and Ana Chac\'on-Tanarro for useful discussions.
PC acknowledges the financial support of the European Research Council (ERC; project PALs 320620).



\begin{appendix}

\section{Map of the $N$($c$-C$_3$H$_2$)/$N$(CH$_3$OH) ratio}
\begin{figure}
 \centering
 \includegraphics [width=0.4\textwidth]{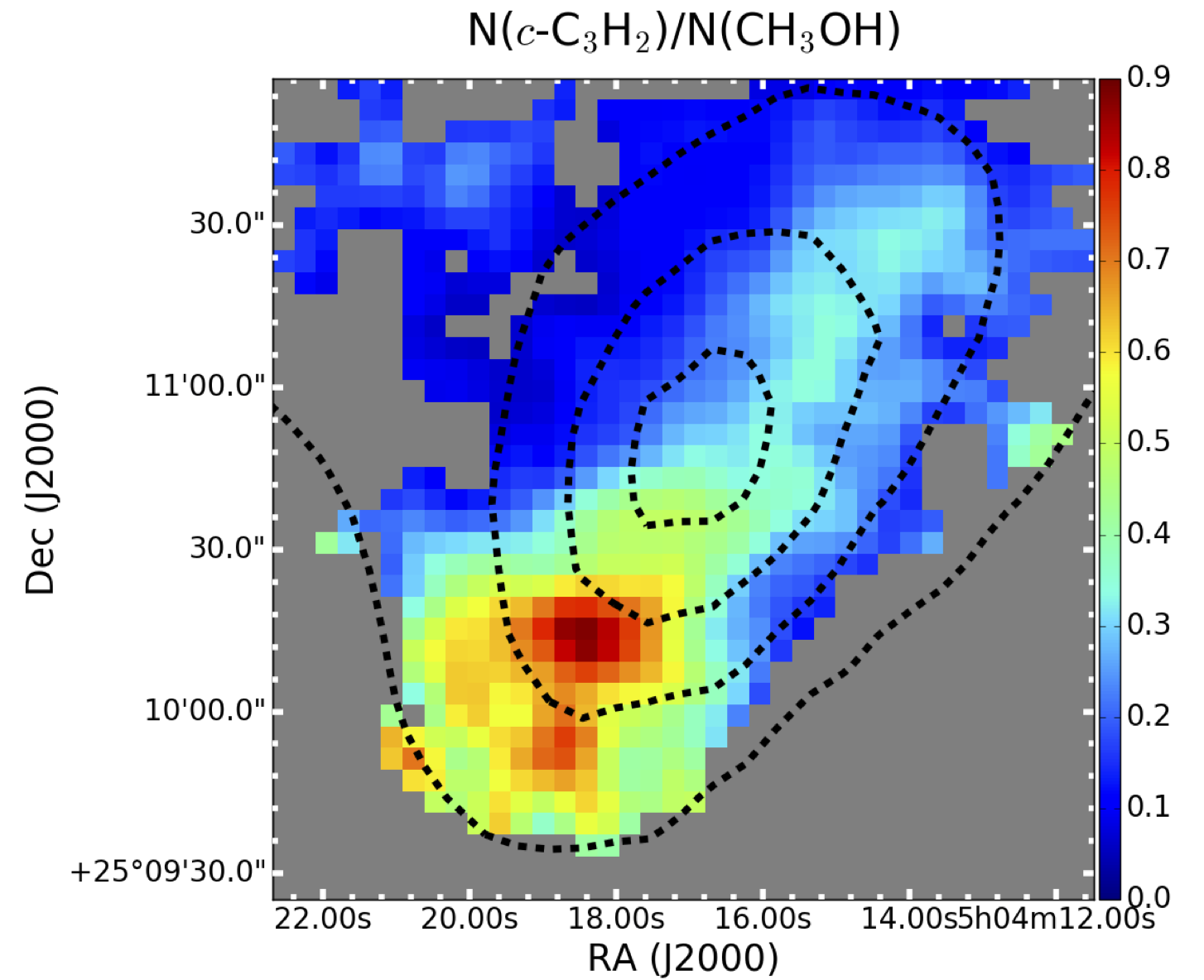}
 \caption{Maps of the total column density ratio of cyclopropenylidene and methanol. The black dashed line represent the 90\%, 70\%, 50\% and 30\% of the H$_2$ column density peak value derived from {\sl Herschel} maps.}
  \label{fig:ratio}
\end{figure}

 Figure \ref{fig:ratio} shows a map of the column density ratio of $c$-C$_3$H$_2$ and CH$_3$OH. Only the pixels where the integrated intensity over the average error is larger than 3 are plotted. While both molecules are depleted towards the center, their abundance ratio is not constant across the core because they present a different distribution. The peak value of the $N$($c$-C$_3$H$_2$)/$N$(CH$_3$OH) ratio is toward the South-East edge of the cloud. 
\section{Kinematics}
\begin{figure}
 \centering
 \includegraphics [width=0.5\textwidth]{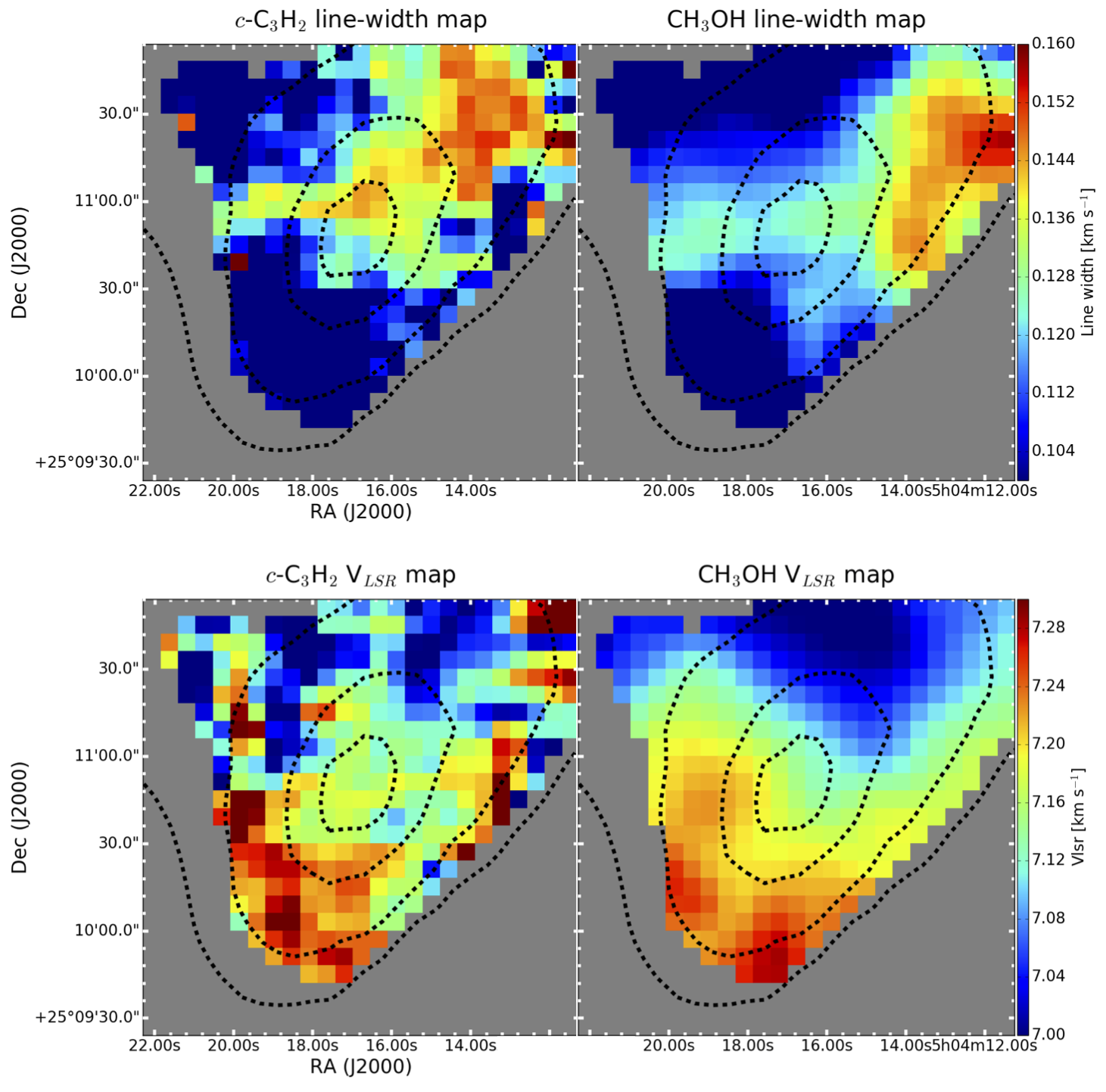}
 \caption{Maps of the line-width (upper panels) and V$_{LSR}$ (lower panels) of cyclopropenylidene and methanol. Both the velocity and line-width structure is similar in the two tracers (see text).}
  \label{fig:vlsr}
\end{figure}

The frequency resolution is high enough to reveal some interesting features on the core kinematics from the line profiles.
The upper panels in Figure \ref{fig:vlsr} show the line-width maps of $c$-C$_3$H$_2$ and CH$_3$OH, while the lower panels show the V$_{LSR}$ maps. The black dashed lines represent  the contours of the H$_2$ column density. Here the line width corresponds to the intensity weighted dispersion of the velocity ($\sigma$).
The line width of both molecules is larger along the major axis of L1544 (South-East to North-West). Besides the well known line width increase measured in N$_2$H$^+$ (1-0) lines toward the centre of the core due to contraction motions \citep{cas02a}, cyclopropenylidene and methanol also show some line broadening toward the North-West and toward the East. This may suggest the presence of a non-uniform turbulent field, possibly caused by non-uniform accretion of envelop material onto the core.
The V$_{LSR}$ maps clearly show that there is a gradient from the north to the South of the core, with the velocity increasing towards the south part of L1544. The line profiles predicted by the model from \cite{cio00} shown in Figure 6 in \cite{cas02a} show the same behaviour, and to some extent also the observed line profiles of N$_2$H$^+$ in Figure 7 of the same paper. Therefore, despite the different distributions of $c$-C$_3$H$_2$ and CH$_3$OH in L1544, the two molecules trace similar kinematic patterns.

\end{appendix}

\end{document}